\documentclass[traditabstract]{aa} % for the abstract without structuration 
\usepackage{epsfig,amsmath}
\usepackage[dvips]{color}
\usepackage{verbatim}
\usepackage{graphicx}
\usepackage{txfonts}
\usepackage{natbib}
\bibpunct{(}{)}{;}{a}{}{,}
\usepackage{hyperref}
\usepackage{rotating}
\definecolor{Black}{named}{Black}
\definecolor{Red}{named}{Red}
\definecolor{Green}{named}{Green}
\definecolor{Blue}{named}{Blue}

\begin{document}

% \title{GRBs \& QSO-DLAs: Is the metals-to-dust ratio a universal constant?}
\title{The metals-to-dust ratio to very low metallicities using GRB and QSO absorbers; extremely rapid dust formation}
\author{Tayyaba Zafar\inst{1}
 \and Darach Watson\inst{2}}

\institute{Aix Marseille Universit\'e, CNRS, LAM (Laboratoire d'Astrophysique de Marseille) UMR 7326, 13388, Marseille, France. 
\and Dark Cosmology Centre, Niels Bohr Institute, University of Copenhagen,
Juliane Maries Vej 30, DK-2100 Copenhagen, Denmark.}

\titlerunning{The metals-to-dust ratios of GRB and QSO absorbers}
\authorrunning{T. Zafar \& D. Watson}

\offprints{tayyaba.zafar@oamp.fr}

\date{Received  / Accepted }

\abstract {Among the key parameters defining the interstellar media (ISM) of galaxies
is the fraction of the metals that are locked up in dust: the metals-to-dust
ratio. This ratio bears not only on the ISM and its evolution, but
also particularly on the origin of cosmic dust. We combine extinction and
abundance data from $\gamma$-ray burst (GRB) afterglows with similar data
from quasar (QSO) foreground absorbers, as well as from multiply-imaged
galaxy-lensed QSOs, to determine the metals-to-dust ratios for
lines of sight through a wide diversity of galaxies from blue, dwarf
starbursts to massive ellipticals, across a vast range of redshifts
$z=0.1-6.3$, and nearly three orders of magnitude of column density and
metal abundance. The GRB and lensed QSO extinction methods are the most
reliable that are available outside the Local Group (LG), allowing absolute extinction
measurements.  We thus determine the metals-to-dust ratio in a unique way,                  
providing direct determinations of in situ gas and dust columns without
recourse to assumptions with large uncertainties. We find that the
metals-to-dust ratios in these systems are surprisingly close to the value  
for the LG, with a mean value of
$10^{21.2}$\,cm$^{-2}\,A_V$\,mag$^{-1}$ and a standard deviation of
0.3\,dex, compared to the Galactic value of
$10^{21.3}$\,cm$^{-2}A_V$\,mag$^{-1}$ (in units of the Galactic gas-to-dust
ratio).  There is no evidence of deviation from this mean ratio as a
function of metallicity, even down to our lowest metallicity of
$0.01\,Z/Z_{\sun}$.  The lack of any obvious dependence of the
metals-to-dust ratio on column density, galaxy type or age, redshift,
or metallicity indicates a close correspondence between the formation of the
metals and the formation of dust.  Any delay between the formation of metals
and dust must be shorter than the typical metal-enrichment times of these             
galaxies, i.e.\ shorter than a few Myr.  Formation of the bulk of the dust 
in low mass stars is therefore ruled out by these data at any cosmic epoch.
Furthermore, dust destruction must not dominate over formation/growth in    
virtually any galaxy environment.  The close correlation between metals and 
dust is a natural consequence of the formation of the bulk of cosmic dust in
supernovae.  Grain growth in the ISM, if it is to be the dominant cosmic  
dust formation mechanism, is strongly constrained by these data to operate
on very short timescales.}

\maketitle{}

%________________________________________________________________
%Introduction
\section{Introduction}

The important constituents of the ISM are gas, metals, and dust which play a
crucial role in the properties of star formation and in the appearance,
formation, and evolution of galaxies. Dust grains are made up of metals,
primarily O, Si, C, Mg, and Fe, which are introduced into the ISM by stars
close to the end, or at the end of their lives.  There is, however, significant
debate over the origin of the bulk of the dust and, related to this, how
quickly dust can form from these metals.  This is an important issue
from a cosmic perspective for a number of reasons.  For example, fully half
of the non-primordial radiation in the universe is reprocessed through dust
grains, and among early bursts of star formation the presence of dust may
have helped cool the gas in the formation of the second generation of stars
\citep{schneider12} and may have provided the catalyst for the formation of
molecular hydrogen, the driver of star formation.  The front-running candidates
for the formation of the bulk of the dust are the
cool, dense envelopes of evolved, low mass stars \citep{gail09}, supernova
(SN) ejecta \citep{dunne03}, or grain growth in the dense ISM
\citep{draine09}. The observed metals-to-dust ratio and its evolution with
total metallicity is a key parameter in tracing the origin of the bulk of
the dust and how it evolves in the ISM. For example, if most of the dust is
formed in SNe and is not largely destroyed thereafter, the metals-to-dust ratio
should be fairly constant over time and be very similar in different
types of galaxies. If, on the other hand, low-mass stellar envelopes form
most of the dust, there should be a delay between the formation of metals
and the formation of dust of a few billion years, which should be detectable
in an apparent diversity and evolution of the metals-to-dust ratio.

In this paper we determine the metals-to-dust ratios for a large, diverse
sample of galaxies as a function of total column density and metallicity. 
We use an absorption and extinction methodology to directly determine the metal and
dust columns along individual lines of sight to produce
metals-to-dust ratios that are far less prone to systematic and modeling
uncertainties than those produced from emission measurements or from
depletion estimates.  Furthermore, our sample covers galaxies up to the edge
of the reionization epoch, in a redshift range from $z=0.1-6.3$, and over an
unprecedented metallicity range of [M/H]$=-2.0$--0.5.

We use extinctions and metal column densities from a large sample of
$\gamma$-ray burst (GRB) afterglows, together with $\ion{H}{i}$ column
densities where available.  We also use gas and metal column densities from
QSO foreground absorption systems together with extinctions derived from
template fitting for these objects, and we use optical/UV extinctions and
total metal column densities derived from X-ray spectroscopy for a few
galaxy-lensed quasars.  We compare these metals-to-dust ratios of GRB
afterglows, QSO-damped Ly$\alpha$ absorbers (DLAs), and lensed galaxies to the value obtained in the Magellanic Clouds and in the Milky Way (MW). 

In Sect. 2 we present our sample selection criteria and describe results from
the analysis.  In Sects. 3 and 4 we provide results and a brief discussion.  The
conclusions are summarized in Sect. 5.  Throughout this paper we use cosmic
abundances from \citet{anders89}.

\begin{figure}
  \centering
{\includegraphics[width=\columnwidth,clip=]{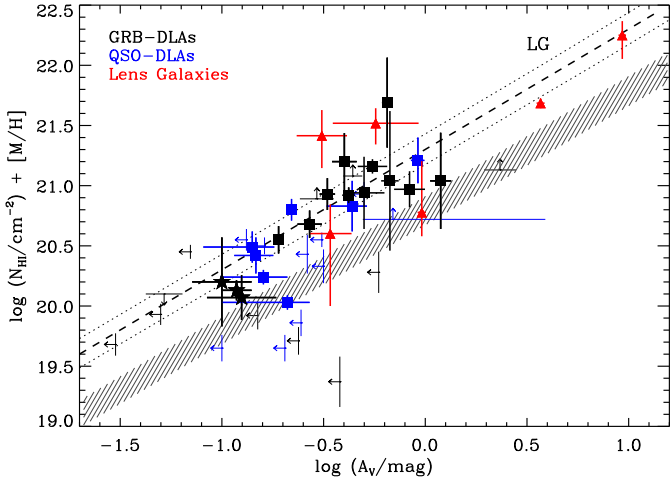}}
     \caption{Equivalent metal column density; $N_{\ion{H}{i}}$ corrected for metallicity in GRB afterglows (black), QSO-DLAs (blue), and nearby lensed galaxies (red) from \citet[][black]{dai09} as a function of dust extinction. The dashed curve represents the metals-to-dust ratio for the LG environments. The dotted lines are metals-to-dust ranges for the MW. The Zn-based metallicities are shown by squares while the stars represent the S-based metallicities. The gray shaded area represents the lower bound below which objects probably cannot exist (see text).}
         \label{metall}
  \end{figure}

%________________________________________________________________
\section{Methods}

Previous studies of the metals-to-dust ratio outside the LG
have relied mostly on emission measurements; from dust, from $\ion{H}{i}$,
and a metallicity estimate from emission line ratios.  These measurements
presented a diversity of results depending on the methodology used and the
assumptions made.  The advantages of these studies are that they offer a
total census of each galaxy to the sensitivity limits of the instrument or
observation, and that the galaxies can often be resolved, allowing
examination of the spatial distribution of ISM components.  The inherent
limitations of emission studies due to sensitivity limitations apply; the
measurements of gas-phase metallicity do not cover the same region as the
dust emission, which does not cover the same area as the 21\,cm $\ion{H}{i}$
emission; and luminous galaxies are more easily surveyed, and the studies can
typically be conducted at low redshift.  Furthermore, conversions from
far-infrared (FIR) emission to dust mass require significant, uncertain
assumptions even where a measurement of the dust temperature can be made.

The ISM of galaxies can also be pursued via absorption of strong
backlighting sources. These sources are particularly useful where their
intrinsic spectra are well known.  And while only one specific sightline is
observed through these galaxies, the advantages of working in absorption are
numerous: a) essentially all stages of cosmic time are available, since some
of these sources can be observed spectroscopically as far back as $z\sim8$
\citep{tanvir}; b) the method is highly sensitive, and even very low
metallicities and dust contents are detectable; c) the observations always
compare the same column of material, whether in dust, gas, or metals, meaning
that there is no bias due to the relative sensitivity to these different
components across a given galaxy; and d) the inherent luminosity of the absorber
plays no role in its detection.  Absorption spectroscopy in the restframe UV
can also place direct limits on the column density of H$_2$ (Kr\"uhler et
al.  in prep; \citealt{prochaska09}; \citealt{noterdaeme08}).  Gamma-ray burst afterglows and QSOs are the obvious backlighting candidates for this technique, as both are extremely
luminous, have quite well-known spectra, and are thus excellent probes of
absorbing gas and extinguishing dust along the line of sight.  Most GRB
afterglows are strongly absorbed by their own host galaxy's ISM
\citep[e.g.][]{vreeswijk04,fynbo09,watson07} and so probe the ISM of
strongly star-forming galaxies \citep{jakobsson12} from $z=0.0086-8.2$,
typically far away from the burst itself
\citep{watson07,vreeswijk04,delia09}.  Quasars, on the other hand, are used as
probes of foreground galaxies along the line of sight, and so typically
intersect the low-surface density outskirts of galaxies
\citep[e.g.][]{chen05a}.  Measurements of the metals-to-dust ratio in QSO
foreground absorbers have been made using the relative depletion of
refractory elements out of the gas phase as a measurement of the dust column
\citep{pei99,vladilo06}, rather than a direct extinction measure.  Finally,
QSOs are also occasionally strongly lensed and multiply imaged by the
gravitational potential of a massive foreground galaxy.  Comparison of the
two lensed images of these systems typically shows one image passing through
the relatively dense ISM of the lensing galaxy, allowing the relative
extinction and absorption properties of the two sightlines to be determined
\citep[e.g.][]{toft00,dai09}.

\subsection{Sample selection and analysis}

In a previous work we produced the largest sample of spectroscopic
extinction curves to date outside the MW, using GRB afterglows
\citep{zafar11}. From this sample we have optical extinction estimates for 9 GRB afterglows.  In
addition, we have taken other GRBs from the literature with optical extinction, metal column
density, and $\ion{H}{i}$ column density measurements, making up a total of 25 GRB afterglows. These additional GRBs were selected because they have their optical extinction derived from X-ray--to--optical/near-infrared spectral energy distribution fitting.  To the GRB
data we have added 17 QSO foreground absorbing systems and 6
gravitationally-lensed, multiply-imaged quasar systems.  These are the ones
we can find in the literature where both metal column density and
reddening/extinction estimates are available (See Table~\ref{sample} for
details).

The extinction of GRBs is estimated from the deviation from a single or
broken power-law of their X-ray--to--optical/near-infrared spectral energy
distribution.  The extinction from lensed quasars is derived from the
difference between the spectra of the multiple images.  These two methods
are the most reliable in determining extinctions at cosmological distances
and can be used to determine absolute extinction curves.  For QSO-DLAs
reddening is measured either from QSO colors or based on template fitting,
and so these values are less robust. 
The reddening template used is the Small Magellanic Cloud extinction curve of
\citet{pei} except for one case, Q\,1157+6135, where a MW-like extinction
curve and a 2175\,\AA\ bump is detected at the redshift of the DLA
\citep{wang12}.  To determine total metallicities and avoid biases related
to depletion of metals from the gas phase onto dust, we derived metal column
densities from non-refractory elements, either Zn or S from low-ionization
lines.  To be consistent throughout, we used solar abundances from
\citet{anders89} for metallicity determination.  \citet{anders89} abundances
(for oxygen in particular) are significantly higher than indicated by direct
measurements of the solar spectrum \citep{asplund09}; however, they are
probably a better estimate of the typical Galactic ISM abundance
\citep[see][]{watson11} and are the metallicities typically used in X-ray
measurements to derive equivalent column densities of hydrogen. For GRBs\,061121, 070506, 080319C, 080413B, 080605, and 080905B Zn equivalent widths from \citet{fynbo09} are used to obtain column densities using the optically thin approximation \citep[see also][]{laskar11}. In order to
be able to compare metal column densities, we derived a total equivalent
column density for the GRB afterglows and QSO-DLAs by modifying
$N_{\ion{H}{i}}$ for metallicity (i.e.\ log ($N_\ion{H}{i}$/cm$^{-2}$) $+$
[M/H]). This then allows us to compare the total metal column density in a
common reference, i.e.\ in terms of the equivalent gas column density for
the Galaxy. Gamma-ray bursts 061121, 080319C 180413A, and 080605 have no $N_{\ion{H}{i}}$ estimates because they are at $z$ $<$ 2. Therefore, their metallicities could not be derived. However, for these GRBs, the total metal column could be derived directly from the S or Zn column densities. For this analysis we also use results from galaxy-lensed quasars from \citet{dai09}, where total metal column densities can be obtained from the soft X-ray absorption something that is not possible for long GRBs because of their anomalous X-ray column densities \citep{watson2013}, again using the same abundances. The metallicities of lensed quasars are not derived because of the lack of $N_{\ion{H}{i}}$ column density estimates.
  
\section{Results}

In Fig.~\ref{metall} we show the optical extinction and equivalent metal
column density for GRB afterglows and QSO-DLAs.  In this comparison, we span
more than three orders of magnitude in column density and include a very
diverse range of objects, everything from massive ellipticals
\citep{toft00}, to the hearts of star-forming galaxies \citep{watson07} and
to the outskirts of galaxies \citep{chen05a,ellison05}.  Furthermore, our
results span very large redshift ($z$=0.1--6.3) and metallicity ranges
([M/H]=$-2.0$--0.5).  We compared all these results to the metals-to-dust
relation for the Galaxy and Magellanic Clouds (here referred to as LG).  For
the LG, we use here a canonical value of metals-to-dust ratio $\approx
2\times 10^{21}$ cm$^{-2}$ mag$^{-1}$ \citep[see][and references therein]{watson11}.

\begin{figure}
  \centering
{\includegraphics[width=\columnwidth,clip=]{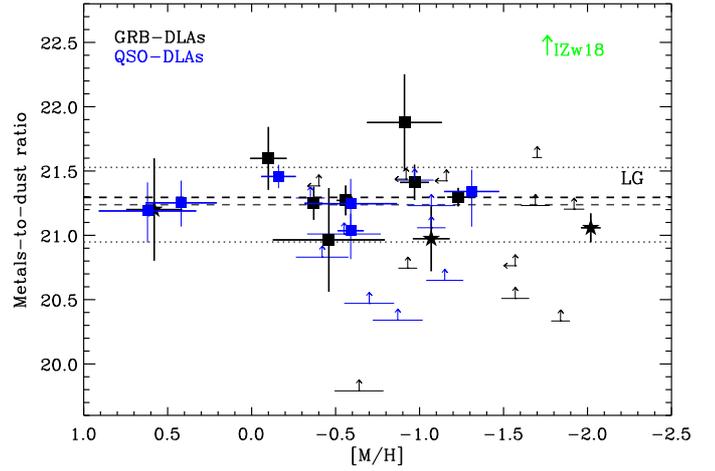}}
     \caption{Metals-to-dust ratio ((log $N_{\ion{H}{i}}$/cm$^{-2}$ $+$ [M/H]) - log $A_V$/mag) versus metallicity for GRB afterglows (black) and QSO-DLAs (blue). The black dashed line represents the LG metals-to-dust relation. The gray dashed line is the mean metals-to-dust ratio of our sample and standard deviation is illustrated by dotted lines. The symbols have the same meaning as in Fig. \ref{metall}. For comparison, the upper limit
              derived by \citet{herrera12} from SED modelling of the lowest
              metallicity galaxy known in the local universe, I\,Zw\,18, is plotted in green.}
         \label{dustmet}
  \end{figure}

Systems significantly below the LG relation are not expected, since a very
large fraction of refractory metals are already in the dust phase in the
Galaxy and Magellanic Clouds \citep{draine03}.  Practically, systems with
less than about half the metals-to-dust ratio of the LG cannot exist, as
there are simply not enough metals available to form more dust. Comparing
the LG metals-to-dust ratio with the mean LG {\it metals-to-dust mass ratio}
\citep[i.e.  2.3,][and references therein]{pei99}, we have
shaded the region in Fig.~\ref{metall} below which objects cannot
exist for this reason. On the other hand, objects that are very metal-rich
and at the same time have relatively low $A_V$ (top-left region of
Fig.~\ref{metall}) should be easy to find if they existed. However,
we find no cases that are significantly dust-poor with respect to the
amount of metals present compared to the LG values. We cover a very wide range in column density, but
we see almost the same metals-to-dust ratio everywhere and centered at the
LG relation. The tightness of this relation across so many different
environments, redshifts, metallicities, and column densities is quite
surprising. For our data, we find a mean metals-to-dust ratio of
$10^{21.2}$\,cm$^{-2}\,A_V^{-1}$ (detections only) for our assumed abundances,
with a standard deviation of 0.3\,dex. Including limits in the
analysis, we find a mean metals-to-dust ratio of $10^{21.1}$\,cm$^{-2}\,A_V^{-1}$,
with a standard deviation of 0.4\,dex. Previously \citet{zafar11} found that GRBs have slightly higher metals-to-dust ratios compared to the LG using UV absorption lines. The UV line metallicities in that case were derived based on the solar abundances of \citet{asplund09}. The metals-to-dust ratio was compared to the LG value based on Galactic X-ray data, which almost invariably uses the solar abundance set of \citet{anders89}, because it is representative of typical Galactic ISM abundances \citep{watson11}. In the present work we use S and Zn based metallicities, but adopt a set of solar abundances to make the comparison consistent between the X-ray derived, LG, metals-to-dust ratio, and the UV line column densities (in this case \citealt{anders89}). We find that the metals-to-dust ratio is consistent with the LG relation.

Foreground absorbers may add some contribution to the observed extinction. 
\citet{menard08} have determined a color excess that decreases strongly with
redshift ($E(B-V)\propto (1+z)^{-1.1}$) for their foreground galaxies. 
Using their results, the correction will typically be $A_V<0.05$ for our
QSO-DLAs and GRBs, decreasing to high redshift, and is thus expected to be
within or close to the level of our uncertainties in most cases. Our most constraining datapoint, GRB\,050730 with [M/H] $=-2.02$, is worth a special note. Because of its especially low extinction ($A_V = 0.12\pm0.02$), it may be vulnerable to intervening absorption. We note that it has some intervening foreground absorbers, at $z=$ 3.564, 2.262, 2.253, and 1.772 \citep{delia07}. Using the relation of \citet{menard08}, we estimate that approximately one third of the observed extinction should be due to known foreground absorbers. This correction would make the data more consistent with the LG relation. As we noted above, the mean line of sight should not be heavily affected by intervening absorbers. However, given the dependence of our result at low metallicities (below [M/H]$<-1.3$) on this single datapoint, one should be cautious about whether foreground absorbers at lower redshifts ($z\lesssim1.2$) could contribute in this specific case, since most of the flux shortward of 6200\,\AA\ is removed by hydrogen Lyman absorption.

%We determined the metals-to-dust ratio in mass terms for our lensed galaxies, GRB and QSO-DLA sample by: $(i)$ using the mass-to-extinction conversion for the Galaxy or the Magellanic Clouds (all $\sim4.4\times10^{-5}$\,g\,cm$^{-2}$\,mag$^{-1}$) and $ii)$ assuming the mass of the gas $2.2\times 10^{-24}$g/H (assuming H, He mass and Galaxy metallicity). This corresponds to a metals-to-dust mass ratio of 96 with a standard deviation of 88.

\subsection{Metals-to-dust ratio versus metallicity}

\begin{figure}
  \centering
{\includegraphics[width=\columnwidth,clip=]{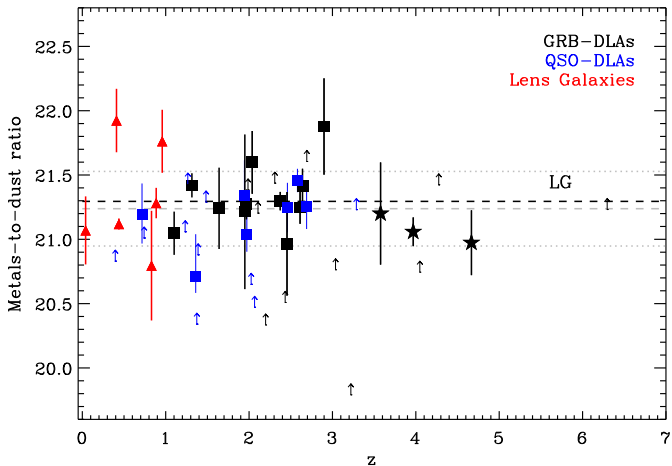}}
     \caption{Metals-to-dust ratio versus redshift for GRB afterglows (black), QSO-DLAs (blue), and lensed galaxies (red). The black dashed line represents the LG metals-to-dust relation. The gray dashed line is the mean metals-to-dust ratio of our sample and its standard deviation is illustrated by dotted lines. The symbols have the same meaning as in Fig. \ref{metall}.}
         \label{metalz}
  \end{figure}

One of the key diagnostics of where and when dust is formed is its variation
with metallicity.  Our GRB (and QSO-DLA) data are rich enough that we can
determine metallicities for our sample above $z\sim1.8$ (due to the
atmospheric window for H\,\textsc{i}\,Ly$\alpha$).  We therefore show how
the metals-to-dust ratio varies as a function of metallicity for these
objects in Fig.~\ref{dustmet}.  There is no evidence of any variation in the
metals-to-dust ratio with metallicity, even objects with metallicities as
low as $\sim1$\% of the solar metallicity show no indication of a different
metals-to-dust ratio. Finally, we show the metals-to-dust ratio as a
function of redshift (Fig.~\ref{metalz}), and again, we find no apparent
change, i.e.\ the metals-to-dust ratio we find at $z\sim6.3$ is the same as
in our own Galaxy.  

\citet{pei99} estimated dust masses by measuring the depletion of [Cr/Zn]
and reported that the mean dust-to-metals ratio was roughly constant over a
redshift range $0\lesssim z \lesssim3$.  In the present study, we are
measuring the full dust column directly through extinction (GRBs) or
reddening (QSOs) rather than inferring its presence from the relative
depletion of Cr.  Our data is consistent with the results found by
\citet{pei99} but extends the redshift range from $z<3$ to $z>6$, and covers
a much wider variety of objects, dealing not only with QSO-DLAs, but also
with the hearts of star-forming galaxies.
  
\section{Discussion} 

The origin of cosmic dust is currently a major issue in astrophysics and
cosmology.  The three candidates for forming the bulk of the dust are
condensation in the envelopes of low mass, evolved stars \citep[AGB dust
formation,][]{gail09}, condensation in SN ejecta \citep{dunne03}, and growth
in dense molecular clouds \citep{draine09}.  While it has long been
advocated on the basis of models of dust formation in AGB stars and dust
destruction in the ISM that AGB stars are not sufficient to explain all of
the dust in the Galaxy \citep[e.g.][]{draine79,gehrz89}, they dominate the
pre-solar grain populations found to date \citep{nittler}, and were
believed to be the dominant form of dust from stars \citep{gail09}. 
Furthermore, it has been proposed that they were major contributors to dust in the
early universe \citep{valiante09,cherchneff09}. Conversely, \citet{maiolino04,dwek07,michalowski10,gall11,gall11b,dwek11} claimed that AGB stars are not efficient enough at high redshifts and that dust is due to SNe or ISM dust grain growth.

For dominant dust formation
by AGB stars, there is a significant time delay between the formation of the
bulk of the metals (which are formed in SNe) and the bulk of the dust,
because of the time required to evolve off the main sequence for these
stars.  This delay will introduce significant scatter in the
metals-to-dust ratio, particularly showing deviations at the lowest
metallicities, where the metals-to-dust ratios should be very high, since
the dust has typically not had time to form.  

For bulk dust formation via accretion of dust onto seed particles in the
dense ISM, the growth rate is strongly related to the metallicity of the
ISM.  Indeed, in most analyses there is a critical metallicity below which
grain growth in molecular clouds is likely to be insignificant
\citep[e.g.][]{zhukovska09,asano11}.  The precise critical metallicity is
dependent on the sticking coefficient for the grains, the survival times of
the molecular clouds, and the injected grain size distribution. 
\citet{zhukovska09} indicate that grain growth in molecular clouds only
becomes dominant above a metallicity of [M/H]$\sim-1$.  Again, this
indicates that if dust formation is dominated by this mechanism, there
should be a fall away from a constant metals-to-dust ratio at low
metallicities.  Where precisely this low metallicity value lies is unclear. 
However, down to [M/H]$=-2$, we see no sign of it in our data
(Fig.~\ref{dustmet}).  If grain growth in the ISM does dominate dust masses,
the process must therefore be much more efficient than assumed in
\citet{zhukovska09}, implying an extremely fast dust growth timescale in
higher metallicity systems.  \citet{draine09} makes a simple, order of
magnitude estimate of the timescale required for dust growth in the ISM. 
According to Eq.~8 of \citet{draine09}, a 1\% metallicity would imply an
accretion timescale of 1\,Gyr, far longer than the typical stellar ages of
GRB hosts \citep{christensen04,savaglio09,jakobsson12}, and therefore
inconsistent with our data.  On the other hand, charged Polycyclic Aromatic Hydrocarbons (PAH) could make the initial growth rate a factor of 50 faster than this \citep{draine09} at least
for the growth of the smallest grain, i.e.\ 20\,Myr, which is consistent
with the typical stellar ages of GRB hosts.  However, we should note that
this is extreme, since it relies on PAH growth, which seems unlikely to
drive the bulk of grain growth.  Future measurements of dust and metal
columns in objects with metallicities significantly below 1\% of the solar
abundance should thus distinguish between ISM grain growth and SN-formation
(see below) as the dominant dust-making scenario.  As a corollary, this
analysis also implies that in solar metallicity systems, the grain accretion
timescales are typically less than 1\,Myr. A similarly short timescale has
been inferred on other grounds based on an analysis of dusty high redshift
galaxies (Watson et al. in prep.; see also \citealt{calura08, pipino11}).

Finally, a constant metals-to-dust ratio is a natural consequence of
formation of the bulk of the dust in core collapse SNe.  In this scenario,
if core-collapse SNe turn most of their refractory metals into dust, as long
as there is not wholesale return of the metals to the gas via destruction of
dust in the ISM, then a constant metals-to-dust ratio is what would be
expected across all redshifts, metallicities, and galaxy types.

%\subsection{Possible selection biases}
%Since quasars are selected due to their optical colors, we may not be able to see dusty (and/or extremely metal rich) systems. In our sample we also have the strongest known QSO-DLAs, where metals-to-dust ratio for one of them is consistent with the LG relation \citep{guimaraes12} while the other lie below the LG relation \citep{noterdaeme12}. A substantial fraction of 25--40\% of GRB afterglows is seen in X-ray but not in optical, termed as `dark burst'. These bursts are missing either because of heavy extinction or high redshift. Although we included the known very dusty events, containing the 2175\,\AA\ dust feature, and they follow the LG relation. If there are objects with large amount of gray extinction they possible could be missing because of being faint. 

\subsection{Comparison with other studies}

We cover approximately 2.5 orders of magnitude in metallicity and find that
the ratio in both metal-rich and metal-poor systems is not only constant,
but also that the mean is consistent with the values found for the Galaxy and
Magellanic Clouds.  We do not find any significant variation in the
metals-to-dust ratios of systems with [M/H] $\lesssim-1.5$.  This hints that
metal-poor and metal-rich galaxies form dust in the same way.
\citet{draine07} show that when considering only the spatial regions over
which the dust is detected, the metals-to-dust ratios appear to be constant
for the Spitzer Infrared Nearby Galaxies Survey (SINGS) galaxies, a result
consistent with what we find here, though subject to greater uncertainties,
and covering a far less diverse sample in both metallicity and cosmic age. 

The most sensitive data in this connection, of course, are very low
metallicity systems, for the reasons mentioned above. Chemical
evolution models considering dust destruction by SNe \citep{hirashita02} or
mass outflows from the galaxy \citep{lisenfeld98} predict that at low
metallicities the gas-to-dust ratio should not scale linearly with
metallicity as it seems to at higher metallicities.  In examining dwarf,
metal-poor galaxies, \citet{galametz11} showed that the addition of submm
data to SED modeling was important and resulted in higher dust masses than
when modeled without submm data, and showed once again a tight correlation
of the dust-to-gas mass ratio with metallicity.  In searching for such low
metallicity systems, emission studies have turned to the most metal-poor
system known in the local Universe, the blue compact dwarf galaxy I\,Zw\,18
\citep{herrera12} with [O/H] $=-1.76$ (using \citet{anders89} abundances). 
For I\,Zw\,18, \citet{herrera12} find a higher gas-to-dust ratio compared to
the available metallicity, suggesting that at low metallicities the dust
fraction may fall.  We plot their suggested metals-to-dust ratio in
Fig.~\ref{dustmet} for comparison to our data, but it is worth noting that
they assume a constant metallicity across the galaxy, and have some
uncertainty in their determination of the total dust mass, since they do not
detect cool dust emission, and since the dust mass is highly dependent on
the assumed dust properties, temperature, and dust distribution.  Using only
the same spatial regions \citep[c.f.][]{draine07}, and one of the dust
models, \citet{herrera12} note that the I\,Zw\,18 dust-to-gas ratio is only
a factor of two below that expected for the metallicity, which is within the
scatter observed in the \citet{draine07} sample.  It is therefore unclear so
far whether emission measurements indicate a significant change in the
metals-to-dust ratio at low metallicity.  Future observations with the
Atacama Large Millimeter/submillimeter Array (ALMA) should help resolve that
debate within the uncertainties of emission diagnostics.  

Other observations of the metals-to-dust ratio or gas-to-dust ratio as a
function of metallicity include measurements using many different techniques
in the Galaxy \citep[see][and references therein]{watson11}, in the
Magellanic Clouds \citep{weingartner01,gordon03,bernard08}, as well as
in nearby spiral galaxies \citep{issa90} and dwarf galaxies
\citep{lisenfeld98}.  Recently \citet{smith12} reported that the dust-to-gas
ratio gradient of the Andromeda (M31) galaxy varies radially, consistent
with its metallicity gradient.  All of these measurements are consistent
with our findings here that indicate a universal metals-to-dust ratio
constant to within a factor of 30--40\% (see also \citealt{dai09}, as well as a recent more expanded sample by \citealt{chen13}, again consistent with our conclusions here).  We thus
argue that the metals-to-dust ratio is constant within this scatter, i.e.\
0.3\,dex, down to 1\% solar metallicity and suggest that the modelling
uncertainties associated with emission measures are such that the current
data are consistent with this conclusion.

Apart from the study of \citet{pei99} mentioned in Sect. 3.1, \citet{noterdaeme08} have used the depletions of a sample of QSO absorbers to investigate the dust-to-metals ratios as a function of metallicity and redshift. They found a correlation between the depletion of metals and the metallicity, using principally the depletion of iron-group elements. Their results indicate that fractionally less Fe is taken up in metals as the metallicity drops. Very recently, \citep{decia13} have demonstrated the same effect in GRB host galaxy absorbers. These results appear to contradict our findings here, and may be at odds with those of the SINGS sample \citep{draine09}, which do not show a consistent rise in the metals-to-dust ratio with metallicity. A possible explanation of this may be that Fe does not dominate the dust mass, is synthesised in largely different locations, and appears to be depleted at a very different rate out of the gas phase, compared to the principal dust constituents, Mg, Si, O, and C \citep{jenkins09}. It is possible therefore that the depletion of Fe is not representative of the total dust column. Strong evidence that the dust column follows the Fe depletion would therefore be valuable. Our estimates of $A_V$ are somewhat dependent on the slopes of the extinction curves for GRBs which are often poorly constrained where we do not have infrared data. Our planned analysis of afterglows observed with X-shooter will be valuable in this respect. Finally, our results here at the lowest metallicity are subject to the possible caveat on foreground absorbers toward GRB\,050730 mentioned in Sect. 3.
  
%________________________________________________________________
%Conclusions
\section{Conclusions}

In this work we presented the metals-to-dust ratios of a sample of GRB
afterglows as a function of their metallicities and redshifts.  We
supplemented this sample with QSO-DLAs, and lensed QSOs.  Our data span 3
orders of magnitude in column density and 2.5 in metallicity.  The redshifts
range from $z=0.1$ to $z=6.3$, and galaxies vary in type from blue,
sub-luminous star-forming galaxies to massive ellipticals.  We found the
metals-to-dust ratio to be consistent with the LG relation and constant
within a factor of 30--40\% regardless of metallicity, galaxy type, total
column density, or redshift.  We found that the metals-to-dust ratio of our
sample is always close to the value for the LG (i.e.\
$10^{21.3}$\,cm$^{-2}\,A_V^{-1}$) for all of these systems, with a
mean value of $10^{21.2}$\,cm$^{-2}\,A_V^{-1}$ and a standard
deviation of 0.3\,dex.  This provides clear evidence of a near-universal
metals-to-dust ratio within this scatter down to very low metallicity and in
the early universe.  We infer from this that dust formation closely follows
metal formation with at most a short time delay ($\lesssim1$\,Myr) between
them.  The most natural interpretation of the data are that core collapse
SNe produce the bulk of both the metals and the dust simultaneously.  The
data essentially rule out low mass stars as the origin of the bulk of the
dust mass at any cosmic epoch, and puts strong constraints on models of dust
growth in the dense interstellar medium.

%________________________________________________________________
%Acknowledgments
\begin{acknowledgements}

This work has been funded within the BINGO! (`history of Baryons: INtergalactic medium/Galaxies cO-evolution') project by the Agence Nationale de la Recherche (ANR) under the allocation ANR-08-BLAN-0316-01. The Dark Cosmology Centre is funded by the Danish National Research Foundation. We are thankful to Jens Hjorth, Anja Andersen, Lars Mattsson, Thomas Kr\"{u}hler, and Annalisa De Cia for helpful comments. We are thankful to the GRB and QSO community for observing these remarkable objects and providing a well sampled data set.
\end{acknowledgements}

\bibliographystyle{aa}
\bibliography{metals.bib}{}

\begin{thebibliography}{84}
\expandafter\ifx\csname natexlab\endcsname\relax\def\natexlab#1{#1}\fi

\bibitem[{{Anders} \& {Grevesse}(1989)}]{anders89}
{Anders}, E. \& {Grevesse}, N. 1989, \gca, 53, 197

\bibitem[{{Asano} {et~al.}(2011){Asano}, {Takeuchi}, {Hirashita}, \&
  {Inoue}}]{asano11}
{Asano}, R.~S., {Takeuchi}, T.~T., {Hirashita}, H., \& {Inoue}, A.~K. 2011, in
  Astronomical Society of the Pacific Conference Series, Vol. 445, Why Galaxies
  Care about AGB Stars II: Shining Examples and Common Inhabitants, ed.
  F.~{Kerschbaum}, T.~{Lebzelter}, \& R.~F. {Wing}, 523

\bibitem[{{Asplund} {et~al.}(2009){Asplund}, {Grevesse}, {Sauval}, \&
  {Scott}}]{asplund09}
{Asplund}, M., {Grevesse}, N., {Sauval}, A.~J., \& {Scott}, P. 2009, \araa, 47,
  481

\bibitem[{{Berger} {et~al.}(2006){Berger}, {Penprase}, {Cenko}, {Kulkarni},
  {Fox}, {Steidel}, \& {Reddy}}]{berger06}
{Berger}, E., {Penprase}, B.~E., {Cenko}, S.~B., {et~al.} 2006, \apj, 642, 979

\bibitem[{{Bernard} {et~al.}(2008){Bernard}, {Reach}, {Paradis}, {Meixner},
  {Paladini}, {Kawamura}, {Onishi}, {Vijh}, {Gordon}, {Indebetouw}, {Hora},
  {Whitney}, {Blum}, {Meade}, {Babler}, {Churchwell}, {Engelbracht}, {For},
  {Misselt}, {Leitherer}, {Cohen}, {Boulanger}, {Frogel}, {Fukui}, {Gallagher},
  {Gorjian}, {Harris}, {Kelly}, {Latter}, {Madden}, {Markwick-Kemper},
  {Mizuno}, {Mizuno}, {Mould}, {Nota}, {Oey}, {Olsen}, {Panagia},
  {Perez-Gonzalez}, {Shibai}, {Sato}, {Smith}, {Staveley-Smith}, {Tielens},
  {Ueta}, {Van Dyk}, {Volk}, {Werner}, \& {Zaritsky}}]{bernard08}
{Bernard}, J.-P., {Reach}, W.~T., {Paradis}, D., {et~al.} 2008, \aj, 136, 919

\bibitem[{{Calura} {et~al.}(2008){Calura}, {Pipino}, \& {Matteucci}}]{calura08}
{Calura}, F., {Pipino}, A., \& {Matteucci}, F. 2008, \aap, 479, 669

\bibitem[{{Chen} {et~al.}(2013){Chen}, {Dai}, {Kochanek}, \&
  {Chartas}}]{chen13}
{Chen}, B., {Dai}, X., {Kochanek}, C.~S., \& {Chartas}, G. 2013,
  arXiv:1306.0008

\bibitem[{{Chen} {et~al.}(2005{\natexlab{a}}){Chen}, {Kennicutt}, \&
  {Rauch}}]{chen05a}
{Chen}, H.-W., {Kennicutt}, Jr., R.~C., \& {Rauch}, M. 2005{\natexlab{a}},
  \apj, 620, 703

\bibitem[{{Chen} {et~al.}(2005{\natexlab{b}}){Chen}, {Prochaska}, {Bloom}, \&
  {Thompson}}]{chen05}
{Chen}, H.-W., {Prochaska}, J.~X., {Bloom}, J.~S., \& {Thompson}, I.~B.
  2005{\natexlab{b}}, \apjl, 634, L25

\bibitem[{{Chen} {et~al.}(2007){Chen}, {Prochaska}, {Ramirez-Ruiz}, {Bloom},
  {Dessauges-Zavadsky}, \& {Foley}}]{chen07}
{Chen}, H.-W., {Prochaska}, J.~X., {Ramirez-Ruiz}, E., {et~al.} 2007, \apj,
  663, 420

\bibitem[{{Cherchneff} \& {Dwek}(2009)}]{cherchneff09}
{Cherchneff}, I. \& {Dwek}, E. 2009, \apj, 703, 642

\bibitem[{{Christensen} {et~al.}(2004){Christensen}, {Hjorth}, \&
  {Gorosabel}}]{christensen04}
{Christensen}, L., {Hjorth}, J., \& {Gorosabel}, J. 2004, \aap, 425, 913

\bibitem[{{Dai} \& {Kochanek}(2009)}]{dai09}
{Dai}, X. \& {Kochanek}, C.~S. 2009, \apj, 692, 677

\bibitem[{{De Cia} {et~al.}(2011){De Cia}, {Jakobsson}, {Bj{\"o}rnsson},
  {Vreeswijk}, {Dhillon}, {Marsh}, {Chapman}, {Fynbo}, {Ledoux}, {Littlefair},
  {Malesani}, {Schulze}, {Smette}, {Zafar}, \& {Gudmundsson}}]{decia11}
{De Cia}, A., {Jakobsson}, P., {Bj{\"o}rnsson}, G., {et~al.} 2011, \mnras, 412,
  2229

\bibitem[{{De Cia} {et~al.}(2013){De Cia}, {Ledoux}, {Savaglio}, {Schady}, \&
  {Vreeswijk}}]{decia13}
{De Cia}, A., {Ledoux}, C., {Savaglio}, S., {Schady}, P., \& {Vreeswijk}, P.~M.
  2013, arXiv:1305.1153

\bibitem[{{D'Elia} {et~al.}(2011){D'Elia}, {Campana}, {Covino}, {D'Avanzo},
  {Piranomonte}, \& {Tagliaferri}}]{delia11}
{D'Elia}, V., {Campana}, S., {Covino}, S., {et~al.} 2011, \mnras, 418, 680

\bibitem[{{D'Elia} {et~al.}(2007){D'Elia}, {Fiore}, {Meurs}, {Chincarini},
  {Melandri}, {Norci}, {Pellizza}, {Perna}, {Piranomonte}, {Sbordone},
  {Stella}, {Tagliaferri}, {Vergani}, {Ward}, {Angelini}, {Antonelli},
  {Burrows}, {Campana}, {Capalbi}, {Cimatti}, {Costa}, {Cusumano}, {Della
  Valle}, {Filliatre}, {Fontana}, {Frontera}, {Fugazza}, {Gehrels}, {Giannini},
  {Giommi}, {Goldoni}, {Guetta}, {Israel}, {Lazzati}, {Malesani}, {Marconi},
  {Mason}, {Mereghetti}, {Mirabel}, {Molinari}, {Moretti}, {Nousek}, {Perri},
  {Piro}, {Stratta}, {Testa}, \& {Vietri}}]{delia07}
{D'Elia}, V., {Fiore}, F., {Meurs}, E.~J.~A., {et~al.} 2007, \aap, 467, 629

\bibitem[{{D'Elia} {et~al.}(2009){D'Elia}, {Fiore}, {Perna}, {Krongold},
  {Covino}, {Fugazza}, {Lazzati}, {Nicastro}, {Antonelli}, {Campana},
  {Chincarini}, {D'Avanzo}, {Della Valle}, {Goldoni}, {Guetta}, {Guidorzi},
  {Meurs}, {Mirabel}, {Molinari}, {Norci}, {Piranomonte}, {Stella}, {Stratta},
  {Tagliaferri}, \& {Ward}}]{delia09}
{D'Elia}, V., {Fiore}, F., {Perna}, R., {et~al.} 2009, \apj, 694, 332

\bibitem[{{D'Elia} {et~al.}(2010){D'Elia}, {Fynbo}, {Covino}, {Goldoni},
  {Jakobsson}, {Matteucci}, {Piranomonte}, {Sollerman}, {Th{\"o}ne}, {Vergani},
  {Vreeswijk}, {Watson}, {Wiersema}, {Zafar}, {de Ugarte Postigo}, {Flores},
  {Hjorth}, {Kaper}, {Levan}, {Malesani}, {Milvang-Jensen}, {Pian},
  {Tagliaferri}, \& {Tanvir}}]{delia10}
{D'Elia}, V., {Fynbo}, J.~P.~U., {Covino}, S., {et~al.} 2010, \aap, 523, A36

\bibitem[{{Draine}(2003)}]{draine03}
{Draine}, B.~T. 2003, \araa, 41, 241

\bibitem[{{Draine}(2009)}]{draine09}
{Draine}, B.~T. 2009, in Astronomical Society of the Pacific Conference Series,
  Vol. 414, Cosmic Dust - Near and Far, ed. T.~{Henning}, E.~{Gr{\"u}n}, \&
  J.~{Steinacker}, 453

\bibitem[{{Draine} {et~al.}(2007){Draine}, {Dale}, {Bendo}, {Gordon}, {Smith},
  {Armus}, {Engelbracht}, {Helou}, {Kennicutt}, {Li}, {Roussel}, {Walter},
  {Calzetti}, {Moustakas}, {Murphy}, {Rieke}, {Bot}, {Hollenbach}, {Sheth}, \&
  {Teplitz}}]{draine07}
{Draine}, B.~T., {Dale}, D.~A., {Bendo}, G., {et~al.} 2007, \apj, 663, 866

\bibitem[{{Draine} \& {Salpeter}(1979)}]{draine79}
{Draine}, B.~T. \& {Salpeter}, E.~E. 1979, \apj, 231, 438

\bibitem[{{Dunne} {et~al.}(2003){Dunne}, {Eales}, {Ivison}, {Morgan}, \&
  {Edmunds}}]{dunne03}
{Dunne}, L., {Eales}, S., {Ivison}, R., {Morgan}, H., \& {Edmunds}, M. 2003,
  \nat, 424, 285

\bibitem[{{Dwek} {et~al.}(2007){Dwek}, {Galliano}, \& {Jones}}]{dwek07}
{Dwek}, E., {Galliano}, F., \& {Jones}, A.~P. 2007, \apj, 662, 927

\bibitem[{{Dwek} {et~al.}(2011){Dwek}, {Staguhn}, {Arendt}, {Capak}, {Kovacs},
  {Benford}, {Fixsen}, {Karim}, {Leclercq}, {Maher}, {Moseley}, {Schinnerer},
  \& {Sharp}}]{dwek11}
{Dwek}, E., {Staguhn}, J.~G., {Arendt}, R.~G., {et~al.} 2011, \apj, 738, 36

\bibitem[{{El{\'{\i}}asd{\'o}ttir} {et~al.}(2009){El{\'{\i}}asd{\'o}ttir},
  {Fynbo}, {Hjorth}, {Ledoux}, {Watson}, {Andersen}, {Malesani}, {Vreeswijk},
  {Prochaska}, {Sollerman}, \& {Jaunsen}}]{eliasdottir09}
{El{\'{\i}}asd{\'o}ttir}, {\'A}., {Fynbo}, J.~P.~U., {Hjorth}, J., {et~al.}
  2009, \apj, 697, 1725

\bibitem[{{Ellison} {et~al.}(2005){Ellison}, {Kewley}, \&
  {Mall{\'e}n-Ornelas}}]{ellison05}
{Ellison}, S.~L., {Kewley}, L.~J., \& {Mall{\'e}n-Ornelas}, G. 2005, \mnras,
  357, 354

\bibitem[{{Fynbo} {et~al.}(2009){Fynbo}, {Jakobsson}, {Prochaska}, {Malesani},
  {Ledoux}, {de Ugarte Postigo}, {Nardini}, {Vreeswijk}, {Wiersema}, {Hjorth},
  {Sollerman}, {Chen}, {Th{\"o}ne}, {Bj{\"o}rnsson}, {Bloom}, {Castro-Tirado},
  {Christensen}, {De Cia}, {Fruchter}, {Gorosabel}, {Graham}, {Jaunsen},
  {Jensen}, {Kann}, {Kouveliotou}, {Levan}, {Maund}, {Masetti},
  {Milvang-Jensen}, {Palazzi}, {Perley}, {Pian}, {Rol}, {Schady}, {Starling},
  {Tanvir}, {Watson}, {Xu}, {Augusteijn}, {Grundahl}, {Telting}, \&
  {Quirion}}]{fynbo09}
{Fynbo}, J.~P.~U., {Jakobsson}, P., {Prochaska}, J.~X., {et~al.} 2009, \apjs,
  185, 526

\bibitem[{{Fynbo} {et~al.}(2011){Fynbo}, {Ledoux}, {Noterdaeme}, {Christensen},
  {M{\o}ller}, {Durgapal}, {Goldoni}, {Kaper}, {Krogager}, {Laursen}, {Maund},
  {Milvang-Jensen}, {Okoshi}, {Rasmussen}, {Thorsen}, {Toft}, \&
  {Zafar}}]{fynbo11}
{Fynbo}, J.~P.~U., {Ledoux}, C., {Noterdaeme}, P., {et~al.} 2011, \mnras, 413,
  2481

\bibitem[{{Gail} {et~al.}(2009){Gail}, {Zhukovska}, {Hoppe}, \&
  {Trieloff}}]{gail09}
{Gail}, H.-P., {Zhukovska}, S.~V., {Hoppe}, P., \& {Trieloff}, M. 2009, \apj,
  698, 1136

\bibitem[{{Galametz} {et~al.}(2011){Galametz}, {Madden}, {Galliano}, {Hony},
  {Bendo}, \& {Sauvage}}]{galametz11}
{Galametz}, M., {Madden}, S.~C., {Galliano}, F., {et~al.} 2011, \aap, 532, A56

\bibitem[{{Gall} {et~al.}(2011{\natexlab{a}}){Gall}, {Andersen}, \&
  {Hjorth}}]{gall11}
{Gall}, C., {Andersen}, A.~C., \& {Hjorth}, J. 2011{\natexlab{a}}, \aap, 528,
  A13

\bibitem[{{Gall} {et~al.}(2011{\natexlab{b}}){Gall}, {Andersen}, \&
  {Hjorth}}]{gall11b}
{Gall}, C., {Andersen}, A.~C., \& {Hjorth}, J. 2011{\natexlab{b}}, \aap, 528,
  A14

\bibitem[{{Gehrz}(1989)}]{gehrz89}
{Gehrz}, R. 1989, in IAU Symposium, Vol. 135, Interstellar Dust, ed. L.~J.
  {Allamandola} \& A.~G.~G.~M. {Tielens}, 445

\bibitem[{{Gordon} {et~al.}(2003){Gordon}, {Clayton}, {Misselt}, {Landolt}, \&
  {Wolff}}]{gordon03}
{Gordon}, K.~D., {Clayton}, G.~C., {Misselt}, K.~A., {Landolt}, A.~U., \&
  {Wolff}, M.~J. 2003, \apj, 594, 279

\bibitem[{{Guimar{\~a}es} {et~al.}(2012){Guimar{\~a}es}, {Noterdaeme},
  {Petitjean}, {Ledoux}, {Srianand}, {L{\'o}pez}, \& {Rahmani}}]{guimaraes12}
{Guimar{\~a}es}, R., {Noterdaeme}, P., {Petitjean}, P., {et~al.} 2012, \aj,
  143, 147

\bibitem[{{Herrera-Camus} {et~al.}(2012){Herrera-Camus}, {Fisher}, {Bolatto},
  {Leroy}, {Walter}, {Gordon}, {Roman-Duval}, {Donaldson}, {Mel{\'e}ndez}, \&
  {Cannon}}]{herrera12}
{Herrera-Camus}, R., {Fisher}, D.~B., {Bolatto}, A.~D., {et~al.} 2012, \apj,
  752, 112

\bibitem[{{Hirashita} {et~al.}(2002){Hirashita}, {Tajiri}, \&
  {Kamaya}}]{hirashita02}
{Hirashita}, H., {Tajiri}, Y.~Y., \& {Kamaya}, H. 2002, \aap, 388, 439

\bibitem[{{Hurkett} {et~al.}(2006){Hurkett}, {Osborne}, {Page}, {Rol}, {Goad},
  {O'Brien}, {Beardmore}, {Godet}, {Burrows}, {Tanvir}, {Levan}, {Zhang},
  {Malesani}, {Hill}, {Kennea}, {Chapman}, {La Parola}, {Perri}, {Romano},
  {Smith}, \& {Gehrels}}]{hurkett06}
{Hurkett}, C.~P., {Osborne}, J.~P., {Page}, K.~L., {et~al.} 2006, \mnras, 368,
  1101

\bibitem[{{Issa} {et~al.}(1990){Issa}, {MacLaren}, \& {Wolfendale}}]{issa90}
{Issa}, M.~R., {MacLaren}, I., \& {Wolfendale}, A.~W. 1990, \aap, 236, 237

\bibitem[{{Jakobsson} {et~al.}(2012){Jakobsson}, {Hjorth}, {Malesani},
  {Chapman}, {Fynbo}, {Tanvir}, {Milvang-Jensen}, {Vreeswijk}, {Letawe}, \&
  {Starling}}]{jakobsson12}
{Jakobsson}, P., {Hjorth}, J., {Malesani}, D., {et~al.} 2012, \apj, 752, 62

\bibitem[{{Jenkins}(2009)}]{jenkins09}
{Jenkins}, E.~B. 2009, \apj, 700, 1299

\bibitem[{{Laskar} {et~al.}(2011){Laskar}, {Berger}, \& {Chary}}]{laskar11}
{Laskar}, T., {Berger}, E., \& {Chary}, R.-R. 2011, \apj, 739, 1

\bibitem[{{Ledoux} {et~al.}(2009){Ledoux}, {Vreeswijk}, {Smette}, {Fox},
  {Petitjean}, {Ellison}, {Fynbo}, \& {Savaglio}}]{ledoux09}
{Ledoux}, C., {Vreeswijk}, P.~M., {Smette}, A., {et~al.} 2009, \aap, 506, 661

\bibitem[{{Lisenfeld} \& {Ferrara}(1998)}]{lisenfeld98}
{Lisenfeld}, U. \& {Ferrara}, A. 1998, \apj, 496, 145

\bibitem[{{Maiolino} {et~al.}(2004){Maiolino}, {Schneider}, {Oliva}, {Bianchi},
  {Ferrara}, {Mannucci}, {Pedani}, \& {Roca Sogorb}}]{maiolino04}
{Maiolino}, R., {Schneider}, R., {Oliva}, E., {et~al.} 2004, \nat, 431, 533

\bibitem[{{M{\'e}nard} {et~al.}(2008){M{\'e}nard}, {Nestor}, {Turnshek},
  {Quider}, {Richards}, {Chelouche}, \& {Rao}}]{menard08}
{M{\'e}nard}, B., {Nestor}, D., {Turnshek}, D., {et~al.} 2008, \mnras, 385,
  1053

\bibitem[{{Micha{\l}owski} {et~al.}(2010){Micha{\l}owski}, {Murphy}, {Hjorth},
  {Watson}, {Gall}, \& {Dunlop}}]{michalowski10}
{Micha{\l}owski}, M.~J., {Murphy}, E.~J., {Hjorth}, J., {et~al.} 2010, \aap,
  522, A15

\bibitem[{{Nittler}(2009)}]{nittler}
{Nittler}, L.~R. 2009, \pasa, 26, 271

\bibitem[{{Noterdaeme} {et~al.}(2008){Noterdaeme}, {Ledoux}, {Petitjean}, \&
  {Srianand}}]{noterdaeme08}
{Noterdaeme}, P., {Ledoux}, C., {Petitjean}, P., \& {Srianand}, R. 2008, \aap,
  481, 327

\bibitem[{{Noterdaeme} {et~al.}(2010){Noterdaeme}, {Petitjean}, {Ledoux},
  {L{\'o}pez}, {Srianand}, \& {Vergani}}]{noterdaeme10}
{Noterdaeme}, P., {Petitjean}, P., {Ledoux}, C., {et~al.} 2010, \aap, 523, A80

\bibitem[{{Pei}(1992)}]{pei}
{Pei}, Y.~C. 1992, \apj, 395, 130

\bibitem[{{Pei} {et~al.}(1999){Pei}, {Fall}, \& {Hauser}}]{pei99}
{Pei}, Y.~C., {Fall}, S.~M., \& {Hauser}, M.~G. 1999, \apj, 522, 604

\bibitem[{{Perley} {et~al.}(2009){Perley}, {Cenko}, {Bloom}, {Chen}, {Butler},
  {Kocevski}, {Prochaska}, {Brodwin}, {Glazebrook}, {Kasliwal}, {Kulkarni},
  {Lopez}, {Ofek}, {Pettini}, {Soderberg}, \& {Starr}}]{perley09}
{Perley}, D.~A., {Cenko}, S.~B., {Bloom}, J.~S., {et~al.} 2009, \aj, 138, 1690

\bibitem[{{Pipino} {et~al.}(2011){Pipino}, {Fan}, {Matteucci}, {Calura},
  {Silva}, {Granato}, \& {Maiolino}}]{pipino11}
{Pipino}, A., {Fan}, X.~L., {Matteucci}, F., {et~al.} 2011, \aap, 525, A61

\bibitem[{{Prochaska} {et~al.}(2009){Prochaska}, {Sheffer}, {Perley}, {Bloom},
  {Lopez}, {Dessauges-Zavadsky}, {Chen}, {Filippenko}, {Ganeshalingam}, {Li},
  {Miller}, \& {Starr}}]{prochaska09}
{Prochaska}, J.~X., {Sheffer}, Y., {Perley}, D.~A., {et~al.} 2009, \apjl, 691,
  L27

\bibitem[{{Savaglio} {et~al.}(2009){Savaglio}, {Glazebrook}, \& {Le
  Borgne}}]{savaglio09}
{Savaglio}, S., {Glazebrook}, K., \& {Le Borgne}, D. 2009, \apj, 691, 182

\bibitem[{{Savaglio} {et~al.}(2012){Savaglio}, {Rau}, {Greiner}, {Kr{\"u}hler},
  {McBreen}, {Hartmann}, {Updike}, {Filgas}, {Klose}, {Afonso}, {Clemens},
  {K{\"u}pc{\"u} Yolda{\c s}}, {Olivares E.}, {Sudilovsky}, \&
  {Szokoly}}]{savaglio12}
{Savaglio}, S., {Rau}, A., {Greiner}, J., {et~al.} 2012, \mnras, 420, 627

\bibitem[{{Schady} {et~al.}(2012){Schady}, {Dwelly}, {Page}, {Kr{\"u}hler},
  {Greiner}, {Oates}, {de Pasquale}, {Nardini}, {Roming}, {Rossi}, \&
  {Still}}]{schady12}
{Schady}, P., {Dwelly}, T., {Page}, M.~J., {et~al.} 2012, \aap, 537, A15

\bibitem[{{Schady} {et~al.}(2011){Schady}, {Savaglio}, {Kr{\"u}hler},
  {Greiner}, \& {Rau}}]{schady11}
{Schady}, P., {Savaglio}, S., {Kr{\"u}hler}, T., {Greiner}, J., \& {Rau}, A.
  2011, \aap, 525, A113

\bibitem[{{Schneider} {et~al.}(2012){Schneider}, {Omukai}, {Bianchi}, \&
  {Valiante}}]{schneider12}
{Schneider}, R., {Omukai}, K., {Bianchi}, S., \& {Valiante}, R. 2012, \mnras,
  419, 1566

\bibitem[{{Shin} {et~al.}(2006){Shin}, {Berger}, {Penprase}, {Fox}, {Price},
  {Kulkarni}, {Soderberg}, {West}, {Cote}, \& {Jordan}}]{shin06}
{Shin}, M.-S., {Berger}, E., {Penprase}, B.~E., {et~al.} 2006, arXiv:0608327

\bibitem[{{Smith} {et~al.}(2012){Smith}, {Eales}, {Gomez}, {Roman-Duval},
  {Fritz}, {Braun}, {Baes}, {Bendo}, {Blommaert}, {Boquien}, {Boselli},
  {Clements}, {Cooray}, {Cortese}, {de Looze}, {Ford}, {Gear}, {Gentile},
  {Gordon}, {Kirk}, {Lebouteiller}, {Madden}, {Mentuch}, {O'Halloran}, {Page},
  {Schulz}, {Spinoglio}, {Verstappen}, {Wilson}, \& {Thilker}}]{smith12}
{Smith}, M.~W.~L., {Eales}, S.~A., {Gomez}, H.~L., {et~al.} 2012, \apj, 756, 40

\bibitem[{{Starling} {et~al.}(2007){Starling}, {Wijers}, {Wiersema}, {Rol},
  {Curran}, {Kouveliotou}, {van der Horst}, \& {Heemskerk}}]{starling07}
{Starling}, R.~L.~C., {Wijers}, R.~A.~M.~J., {Wiersema}, K., {et~al.} 2007,
  \apj, 661, 787

\bibitem[{{Tanvir} {et~al.}(2009){Tanvir}, {Fox}, {Levan}, {Berger},
  {Wiersema}, {Fynbo}, {Cucchiara}, {Kr{\"u}hler}, {Gehrels}, {Bloom},
  {Greiner}, {Evans}, {Rol}, {Olivares}, {Hjorth}, {Jakobsson}, {Farihi},
  {Willingale}, {Starling}, {Cenko}, {Perley}, {Maund}, {Duke}, {Wijers},
  {Adamson}, {Allan}, {Bremer}, {Burrows}, {Castro-Tirado}, {Cavanagh}, {de
  Ugarte Postigo}, {Dopita}, {Fatkhullin}, {Fruchter}, {Foley}, {Gorosabel},
  {Kennea}, {Kerr}, {Klose}, {Krimm}, {Komarova}, {Kulkarni}, {Moskvitin},
  {Mundell}, {Naylor}, {Page}, {Penprase}, {Perri}, {Podsiadlowski}, {Roth},
  {Rutledge}, {Sakamoto}, {Schady}, {Schmidt}, {Soderberg}, {Sollerman},
  {Stephens}, {Stratta}, {Ukwatta}, {Watson}, {Westra}, {Wold}, \&
  {Wolf}}]{tanvir}
{Tanvir}, N.~R., {Fox}, D.~B., {Levan}, A.~J., {et~al.} 2009, \nat, 461, 1254

\bibitem[{{Th{\"o}ne} {et~al.}(2013){Th{\"o}ne}, {Fynbo}, {Goldoni}, {de
  Ugarte}, {Campana}, {Vergani}, {Covino}, {Kr{\"u}hler}, {Kaper}, {Tanvir},
  {Zafar}, {D'Elia}, {Gorosabel}, {Greiner}, {Groot}, {Hammer}, {Jakobsson},
  {Klose}, {Levan}, {Milvang-Jensen}, {Nicuesa}, {Palazzi}, {Piranomonte},
  {Tagliaferri}, {Watson}, {Wiersema}, \& {Wijers}}]{thoene12}
{Th{\"o}ne}, C.~C., {Fynbo}, J.~P.~U., {Goldoni}, P., {et~al.} 2013, \mnras,
  428, 3590

\bibitem[{{Th{\"o}ne} {et~al.}(2010){Th{\"o}ne}, {Kann}, {J{\'o}hannesson},
  {Selj}, {Jaunsen}, {Fynbo}, {Akerlof}, {Baliyan}, {Bartolini}, {Bikmaev},
  {Bloom}, {Burenin}, {Cobb}, {Covino}, {Curran}, {Dahle}, {Ferrero}, {Foley},
  {French}, {Fruchter}, {Ganesh}, {Graham}, {Greco}, {Guarnieri}, {Hanlon},
  {Hjorth}, {Ibrahimov}, {Israel}, {Jakobsson}, {Jel{\'{\i}}nek}, {Jensen},
  {J{\o}rgensen}, {Khamitov}, {Koch}, {Levan}, {Malesani}, {Masetti}, {Meehan},
  {Melady}, {Nanni}, {N{\"a}r{\"a}nen}, {Pakstiene}, {Pavlinsky}, {Perley},
  {Piccioni}, {Pizzichini}, {Pozanenko}, {Roming}, {Rujopakarn}, {Rumyantsev},
  {Rykoff}, {Sharapov}, {Starr}, {Sunyaev}, {Swan}, {Tanvir}, {Terra}, {de
  Ugarte Postigo}, {Vreeswijk}, {Wilson}, {Yost}, \& {Yuan}}]{thoene10}
{Th{\"o}ne}, C.~C., {Kann}, D.~A., {J{\'o}hannesson}, G., {et~al.} 2010, \aap,
  523, A70

\bibitem[{{Th{\"o}ne} {et~al.}(2008){Th{\"o}ne}, {Wiersema}, {Ledoux},
  {Starling}, {de Ugarte Postigo}, {Levan}, {Fynbo}, {Curran}, {Gorosabel},
  {van der Horst}, {Llorente}, {Rol}, {Tanvir}, {Vreeswijk}, {Wijers}, \&
  {Kewley}}]{thoene08}
{Th{\"o}ne}, C.~C., {Wiersema}, K., {Ledoux}, C., {et~al.} 2008, \aap, 489, 37

\bibitem[{{Toft} {et~al.}(2000){Toft}, {Hjorth}, \& {Burud}}]{toft00}
{Toft}, S., {Hjorth}, J., \& {Burud}, I. 2000, \aap, 357, 115

\bibitem[{{Valiante} {et~al.}(2009){Valiante}, {Schneider}, {Bianchi}, \&
  {Andersen}}]{valiante09}
{Valiante}, R., {Schneider}, R., {Bianchi}, S., \& {Andersen}, A.~C. 2009,
  \mnras, 397, 1661

\bibitem[{{Vladilo} {et~al.}(2006){Vladilo}, {Centuri{\'o}n}, {Levshakov},
  {P{\'e}roux}, {Khare}, {Kulkarni}, \& {York}}]{vladilo06}
{Vladilo}, G., {Centuri{\'o}n}, M., {Levshakov}, S.~A., {et~al.} 2006, \aap,
  454, 151

\bibitem[{{Vreeswijk} {et~al.}(2004){Vreeswijk}, {Ellison}, {Ledoux}, {Wijers},
  {Fynbo}, {M{\o}ller}, {Henden}, {Hjorth}, {Masi}, {Rol}, {Jensen}, {Tanvir},
  {Levan}, {Castro Cer{\'o}n}, {Gorosabel}, {Castro-Tirado}, {Fruchter},
  {Kouveliotou}, {Burud}, {Rhoads}, {Masetti}, {Palazzi}, {Pian}, {Pedersen},
  {Kaper}, {Gilmore}, {Kilmartin}, {Buckle}, {Seigar}, {Hartmann}, {Lindsay},
  \& {van den Heuvel}}]{vreeswijk04}
{Vreeswijk}, P.~M., {Ellison}, S.~L., {Ledoux}, C., {et~al.} 2004, \aap, 419,
  927

\bibitem[{{Wang} {et~al.}(2012){Wang}, {Zhou}, {Ge}, {Jiang}, {Lu},
  {Prochaska}, {Hamann}, {Wang}, {Wang}, \& {Yuan}}]{wang12}
{Wang}, J.-G., {Zhou}, H.-Y., {Ge}, J., {et~al.} 2012, \apj, 760, 42

\bibitem[{{Watson}(2011)}]{watson11}
{Watson}, D. 2011, \aap, 533, A16

\bibitem[{{Watson} {et~al.}(2006){Watson}, {Fynbo}, {Ledoux}, {Vreeswijk},
  {Hjorth}, {Smette}, {Andersen}, {Aoki}, {Augusteijn}, {Beardmore}, {Bersier},
  {Castro Cer{\'o}n}, {D'Avanzo}, {Diaz-Fraile}, {Gorosabel}, {Hirst},
  {Jakobsson}, {Jensen}, {Kawai}, {Kosugi}, {Laursen}, {Levan}, {Masegosa},
  {N{\"a}r{\"a}nen}, {Page}, {Pedersen}, {Pozanenko}, {Reeves}, {Rumyantsev},
  {Shahbaz}, {Sharapov}, {Sollerman}, {Starling}, {Tanvir}, {Torstensson}, \&
  {Wiersema}}]{watson06}
{Watson}, D., {Fynbo}, J.~P.~U., {Ledoux}, C., {et~al.} 2006, \apj, 652, 1011

\bibitem[{{Watson} {et~al.}(2007){Watson}, {Hjorth}, {Fynbo}, {Jakobsson},
  {Foley}, {Sollerman}, \& {Wijers}}]{watson07}
{Watson}, D., {Hjorth}, J., {Fynbo}, J.~P.~U., {et~al.} 2007, \apjl, 660, L101

\bibitem[{{Watson} {et~al.}(2013){Watson}, {Zafar}, {Andersen}, {Fynbo},
  {Gorosabel}, {Hjorth}, {Jakobsson}, {Kr{\"u}hler}, {Laursen}, {Leloudas}, \&
  {Malesani}}]{watson2013}
{Watson}, D., {Zafar}, T., {Andersen}, A.~C., {et~al.} 2013, \apj, 768, 23

\bibitem[{{Weingartner} \& {Draine}(2001)}]{weingartner01}
{Weingartner}, J.~C. \& {Draine}, B.~T. 2001, \apj, 548, 296

\bibitem[{{Zafar} {et~al.}(2012){Zafar}, {Watson}, {El{\'{\i}}asd{\'o}ttir},
  {Fynbo}, {Kr{\"u}hler}, {Schady}, {Leloudas}, {Jakobsson}, {Th{\"o}ne},
  {Perley}, {Morgan}, {Bloom}, \& {Greiner}}]{zafar12}
{Zafar}, T., {Watson}, D., {El{\'{\i}}asd{\'o}ttir}, {\'A}., {et~al.} 2012,
  \apj, 753, 82

\bibitem[{{Zafar} {et~al.}(2011{\natexlab{a}}){Zafar}, {Watson}, {Fynbo},
  {Malesani}, {Jakobsson}, \& {de Ugarte Postigo}}]{zafar11}
{Zafar}, T., {Watson}, D., {Fynbo}, J.~P.~U., {et~al.} 2011{\natexlab{a}},
  \aap, 532, A143

\bibitem[{{Zafar} {et~al.}(2010){Zafar}, {Watson}, {Malesani}, {Vreeswijk},
  {Fynbo}, {Hjorth}, {Levan}, \& {Micha{\l}owski}}]{zafar10}
{Zafar}, T., {Watson}, D.~J., {Malesani}, D., {et~al.} 2010, \aap, 515, A94

\bibitem[{{Zafar} {et~al.}(2011{\natexlab{b}}){Zafar}, {Watson}, {Tanvir},
  {Fynbo}, {Starling}, \& {Levan}}]{zafar11b}
{Zafar}, T., {Watson}, D.~J., {Tanvir}, N.~R., {et~al.} 2011{\natexlab{b}},
  \apj, 735, 2

\bibitem[{{Zhukovska} \& {Gail}(2009)}]{zhukovska09}
{Zhukovska}, S. \& {Gail}, H.-P. 2009, in Astronomical Society of the Pacific
  Conference Series, Vol. 414, Cosmic Dust - Near and Far, ed. T.~{Henning},
  E.~{Gr{\"u}n}, \& J.~{Steinacker}, 199

\end{thebibliography}
%============== SAMPLE LOG =======================
\begin{table*}
\begin{minipage}[t]{\columnwidth}
\caption{Basic data used to estimate metals-to-dust ratio. The details are provided in columns as (1) GRB name, (2) log $N_{\ion{H}{i}}$, (3) metal column densities, (4) reference element either X=Zn or S or soft X-ray, (5) optical extinction, (6) redshift, and (7) references. 2$\sigma$ upper limits are provided for $A_V$ non-detections.}      
\label{sample} 
\centering
\renewcommand{\footnoterule}{}  % to avoid a line before footnotes     
\setlength{\tabcolsep}{2pt}
\begin{tabular*}{\columnwidth}{@{\extracolsep{\fill}}l c c c c c c}\hline\hline                       
GRB & log $N_{\ion{H}{i}}$ & log $N_X$ & X & $A_V$ & $z$ & Ref.\\
	& cm$^{-2}$ 	& cm$^{-2}$	&		& mag & \\ 
\hline
000926 & $21.30\pm0.21$ &$13.82\pm0.05$ & Zn & $0.38\pm0.05$ & 2.038 & \citet{chen07}, \citet{starling07} \\
030226	& $20.50\pm0.30$  & $<12.70$ & Zn & $0.05\pm0.01$ & 1.987 & \citet{shin06}, \citet{schady11} \\
%030323 & $21.90\pm0.07$ & $15.30\pm0.08$	 & S & $<0.12$ & 3.372 & \citet{vreeswijk04} \\
050401 & $22.60\pm0.30$ & $14.30\pm0.30$ & Zn & $0.65\pm0.04$ & 2.899 & \citet{watson06}, \citet{zafar11} \\
050505 & $22.05\pm0.10$ & $>16.10$ & S & $0.30\pm	0.1$ & 4.275 & \citet{berger06}, \citet{hurkett06} \\
050730 & $22.15\pm0.06$ & $15.34\pm0.10$ & S & $0.12\pm0.02$ & 3.969 & \citet{chen05}, \citet{zafar11} \\
050820A	& $21.05\pm0.10$ & $13.28\pm0.04$ & Zn & $0.27\pm0.04$ & 2.612 & \citet{ledoux09}, \citet{schady12} \\
050904 & $21.62\pm0.02$ & $15.14\pm0.17$ & S & $<0.05$ & 6.295 & \citet{thoene12}, \citet{zafar10,zafar11b} \\
050922C	 & $21.55\pm0.10$ & $14.92\pm0.05$ & S & $<0.24$ & 2.198 & \citet{schady12} \\
060206 & $20.85\pm0.10$ & $15.13\pm0.05$ & S & $<0.23$ & 4.048 & \citet{thoene08}, \citet{schady12} \\
060526 & $20.00\pm0.15$ & $14.58\pm0.25$ & S & $<0.39$ & 3.221 & \citet{thoene10}, \citet{schady11} \\
%060707 & $21.00\pm0.20$ & $>16.28$ & S & $0.08\pm0.02$ & \citet{fynbo09}, \citet{zafar11} \\
061121 & $\cdots$ & $13.76\pm0.06$ & Zn & $0.55\pm0.10$ & 1.315 & \citet{schady12} \\
070506 & $22.00\pm0.30$ &  $>13.68$ & Zn & $0.44\pm0.05$ & 2.308 & \citet{fynbo09}, \citet{zafar11} \\
070802 & $21.50\pm0.20$ & $13.60\pm0.60$	 & Zn & $1.19\pm0.15$ & 2.455 & \citet{eliasdottir09}, \citet{zafar11} \\
071031 & $22.15\pm0.05$ &  $13.05\pm0.03$ & Zn & $<0.07	$ & 2.692 & \citet{ledoux09}, \citet{zafar11} \\
080210 & $21.90\pm0.10$ & $13.53\pm0.14$ & Zn & $0.33\pm0.03$ & 2.641 & \citet{decia11}, \citet{zafar11} \\
080319C & $\cdots$ & $13.64\pm0.60$ & Zn & $0.67\pm0.07$ & 1.949 & \citet{perley09} \\
%080330 & $\cdots$ & $12.79\pm0.06$ & Zn & $<0.02$ & 1.512 & \citet{delia09a,guidorzi09} \\
080413A	 & $21.85\pm0.15$ & $12.88\pm0.07$ & Zn &	$<0.59$ & 2.433 & \citet{ledoux09}, \citet{schady11} \\
080413B & $\cdots$ & $13.57\pm0.15$ & Zn & $0.84\pm0.16$ & 1.101 & \citet{schady12} \\
080605 & $\cdots$ & $13.54\pm0.30$ & Zn & $0.50\pm0.13$ & 1.640 & \citet{zafar12} \\
080607 & $22.70\pm0.15$ & $>16.34$ & S & $2.33^{+0.46}_{-0.43}$ & 3.037 & \citet{prochaska09}, \citet{zafar11} \\
080905B	 & $<22.15$ & $13.52\pm0.13$ & Zn &  $0.42\pm0.03$ & 2.374 & \citet{fynbo09}, \citet{zafar11} \\
081008 & $21.11\pm0.10$ & $13.15\pm0.04$ & Zn & $\approx0.19$ & 1.968 & \citet{delia11} \\
%090205 & $20.73\pm0.05$ & $>15.30$ & S & $<0.10$ & \citet{davanzo10} \\
090323 & $19.62\pm0.33$ & $15.41\pm0.04$ & S & $0.10\pm0.04$ & 3.577 & \citet{savaglio12}, \citet{schady11} \\
090926A & $21.60\pm0.07$ & $14.89\pm0.10$ & S & $<0.03$ & 2.107 & \citet{delia10} \\
100219A & $21.14\pm0.15$ & $15.25\pm0.15$ & S & $0.13\pm0.05$ & 4.667 & \citet{thoene12} \\
\hline
QSO & log $N_{\ion{H}{i}}$ & log $N_X$ & X & $A_V$ & $z$ & Ref. \\
\hline
0013+0004 & $20.80\pm0.01$ & $12.25\pm0.05$ & Zn & $<0.10$ & 2.025 & \citet{vladilo06} \\
0016-0012 & $20.83\pm0.05$ & $12.82\pm0.04$  & Zn & $0.16^{+0.04}_{-0.06}$ &1.973 &  \citet{vladilo06} \\
0121+0027 & $\cdots$ & $>13.32$ & Zn & $0.69^{+3.2}_{-0.2}$ & 1.388 & \citet{vladilo06} \\
0816$+$1446 & $22.00\pm0.10$ & $13.53\pm0.01$ & Zn & $<0.50$ & 3.287 & \citet{guimaraes12} \\
0918$+$1636 & $20.96\pm0.05$ & $13.40\pm0.01$ & Zn & $\approx$0.21 & 2.580 & \citet{fynbo11} \\
0938+4128 & $20.52\pm0.10$ & $12.25\pm0.05$ & Zn & $<0.20$ & 1.373 & \citet{vladilo06} \\
0948+4323 & $21.62\pm0.06$ & $13.15\pm0.01$ & Zn & $<0.31$ & 1.233 & \citet{vladilo06} \\
1010+0003 & $21.52\pm0.07$ & $13.15\pm0.06$ & Zn & $<0.13$ & 1.265 & \citet{vladilo06} \\
1107+0048 & $20.98\pm0.15$ & $13.03\pm0.05$ & Zn & $<0.26$ & 0.741 & \citet{vladilo06} \\
%1135-0010 & $22.10\pm0.05$ & $13.60\pm0.01$ & Zn &  $\approx$0.11 & \citet{noterdaeme12} \\
1157$+$6135 & $21.80\pm0.20$ & $\approx13.80$ & Zn & $0.92\pm0.07$ & 2.459 & \citet{wang12} \\
1159+0112 & $21.80\pm0.10$ & $13.09\pm0.08$ & Zn & $0.14^{+0.04}_{-0.06}$ & 1.944 & \citet{vladilo06} \\
1232-0224 & $20.75\pm0.07$ & $12.93\pm0.12$ & Zn & $<0.32$ & 0.395 & \citet{vladilo06} \\
1237+0647 & $20.00\pm0.15$ & $13.02\pm0.02$ & Zn & $0.15\pm0.03$ & 2.690 & \citet{noterdaeme10} \\
1323-0021 & $20.21\pm0.20$ & $13.43\pm0.05$ & Zn & $0.44^{+0.08}_{-0.11}$ & 0.716 & \citet{vladilo06} \\
1501+0019 & $20.85\pm0.05$ & $13.10\pm0.05$ & Zn & $<0.16$ & 1.483 & \citet{vladilo06} \\
2234+0000 & $20.56\pm0.10$ & $12.46\pm0.02$ & Zn & $<0.25$ & 2.066 & \citet{vladilo06} \\
2340-0053 & $\cdots$ & $12.62\pm0.05$ & Zn & $0.21^{+0.06}_{-0.11}$ & 1.360 & \citet{vladilo06} \\	
\hline
Lensed QSO & log $N_{\ion{H}{i}}$ & log $N_X$ & X-ray & $A_V\footnote{For $R_V=3.1$}$ & $z$ & Ref. \\
\hline
SBS\,0909+523 & $\cdots$ & $20.78^{+0.42}_{-0.20}$ & X-ray & $0.99\pm0.03$ & 0.830 & \citet{dai09} \\ 
B\,1152+199 & $\cdots$ & $21.68^{+0.04}_{-0.04}$ & X-ray & $3.72\pm0.16$ & 0.439 & \citet{dai09} \\
MG\,0414+0534 & $\cdots$ & $21.52^{+0.12}_{-0.18}$ & X-ray & $0.56\pm0.34$ & 0.958 & \citet{dai09} \\
B\,1600+434 & $\cdots$ & $21.42^{+0.21}_{-0.27}$ & X-ray & $0.31\pm0.09$ & 0.410 & \citet{dai09} \\
PKS\,1830-211 & $\cdots$ & $22.25^{+0.12}_{-0.20}$ & X-ray & $9.30\pm0.40$ & 0.886 & \citet{dai09} \\ 
Q\,2237+0305 & $\cdots$ & $20.60^{+0.24}_{-0.60}$ & X-ray & $0.34\pm0.09$ & 0.040 & \citet{dai09} \\
\hline
\end{tabular*}
\end{minipage}
\end{table*}
%==================================================

\end{document}